\documentclass[eqsecnum,showpacs,showkeys,floats,aps,nofootinbib,preprint]{revtex4}
\usepackage{bm,amsfonts, mathtools}
\usepackage{graphicx,color, wasysym}
\usepackage{textcomp}
\usepackage{amsmath,amssymb,latexsym,epsfig}
\def\de{\mathrm{d}}

\def\nn{\nonumber}
\begin{document}

\makeatletter

\title{The Dirac factorization method and the harmonic oscillator}

\author{D. Babusci}
\email{danilo.babusci@lnf.infn.it}
\affiliation{INFN - Laboratori Nazionali di Frascati, via E. Fermi, 40, IT 00044 Frascati (Roma), Italy}

\author{G. Dattoli}
\email{dattoli@frascati.enea.it}
\affiliation{ENEA - Centro Ricerche Frascati, via E. Fermi, 45, IT 00044 Frascati (Roma), Italy}

\begin{abstract}
We apply the Dirac factorization method to the nonrelativistic harmonic oscillator and, more in general, to 
Hamiltonians with a generic potential. It is shown that this procedure naturally leads to a supersymmetric formulation 
of the problems under study. It is also speculated on the physical meaning underlying this method and it is suggested 
that the vacuum field fluctuations can be viewed as the spontaneous emission of the associated two-level system, 
whose quantization is due to the noncommuting nature of the harmonic oscillator canonical variables. 
\end{abstract}

\maketitle


\section{Introduction}\label{s:intro}
As stressed in Ref. \cite{Cohen},  spin systems (i.e., any system with only two energy levels) and harmonic oscillators 
comprise two archetypes in quantum mechanics. Recent experiments \cite{Hofheinz} are going deeper in their peculiar 
nature and in this paper we show that two-level spin systems and oscillator states are naturally entangled in a 
supersymmetric framework. It will be shown, starting from fairly simple mathematical considerations, that such a 
realization involves nothing but that the Dirac factorization method. 

The Hamiltonian of a harmonic oscillator can be written in terms of creation-annihilation operators according to the 
identity\footnote{In the following we use the natural units $\hbar = c = 1$ and assume, for semplicity, $m = 1$ and 
$\omega = 1$.}
\begin{equation}
\label{eq:hoham}
H = \frac12\,(p^2 + q^2) = a^+\,a^- + \frac12
\end{equation}
where
\begin{equation}
a^\pm = \frac1{\sqrt{2}}\,(q \mp i\,p)  
\end{equation}
and
\begin{equation}
[q, p] = i \qquad\qquad [a^-, a^+] = 1\,.
\end{equation}
The Pauli matrices \cite{Cohen} 
\begin{equation}
\sigma_1 = \left(
\begin{array}{rr}
  0 & 1   \\
  1 & 0 
\end{array}
\right) \qquad 
\sigma_2 = \left(
\begin{array}{rr}
  0 & - i   \\
  i & 0 
\end{array}
\right) \qquad 
\sigma_3 = \left(
\begin{array}{rr}
  1 & 0   \\
  0 & - 1 
\end{array}
\right)
\end{equation}
are the realization of the generators of a Clifford algebra and satisfy the identities
\begin{equation}
\label{eq:pauli}
[\sigma_j, \sigma_k] = 2\,i\,\epsilon_{jkm}\,\sigma_m\,, \qquad\qquad
\{\sigma_j, \sigma_k\} = 2\,\delta_{jk}\,,
\end{equation}
which allow us to rewrite the sum of the squares of two operators as follows 
\begin{equation}
\label{eq:sumsq}
A^2 + B^2 = (A\,\sigma_j + B\,\sigma_k)^2 - i\,\epsilon_{jkl}\,[A, B]\,\sigma_l \qquad\qquad (j \neq k)\,.  
\end{equation}
This identity will be referred as \emph{Dirac factorization method}. It was the breakthrough paving the way to the 
relativistic Dirac equation \cite{Dirac}. More recently \cite{PRA}, an analogous procedure has been applied to get 
a unified view of the theory of relativistic wave equations, using identity \eqref{eq:sumsq} as a tool to provide 
a different formulation of the definition of fractional operators and derivatives. We will exploit and further expand the 
formalism developed in Ref. \cite{PRA} by applying the Dirac factorization method to obtain alternative forms of the 
harmonic oscillator Hamiltonian. This technique leads to the introduction of a set of operators which are not the 
ordinary creation-annihilation pairs and are recognized as supercharges. The proposed procedure brings naturally 
to a supersymmetric formulation of the harmonic oscillator Hamiltonian and it is easily generalized to more complicated 
Hamiltonian forms. The method suggests that oscillator-like Hamiltonians are equipped, through Dirac factorization 
method, with a two-level structure providing a fairly transparent understanding of the physical role played by the algebraic 
grading. 

\section{Dirac factorization and the harmonic oscillator}\label{s:dirac}
To proceed in the Dirac factorization of the Hamiltonian \eqref{eq:hoham}, let us introduce the following combination 
of matrices and differential operators 
\begin{equation}
\label{eq:Sigma}
\Sigma = \frac1{\sqrt{2}}\,(q\,\sigma_1 - p\,\sigma_2) = \left(
\begin{array}{cc}
       0  &  a^-  \\
   a^+  & 0   
\end{array}
\right)
\end{equation}
that, according to eqs. \eqref{eq:pauli}, and as a consequence of the non-commuting nature of the operators $q$ and $p$, 
yields 
\begin{equation}
\Sigma^2 = \frac12\,\left\{(q^2 + p^2) - i\,\sigma_1\,\sigma_2\right\}\,,
\end{equation}
and, thus, the Hamiltonian \eqref{eq:hoham} can be rewritten as
\begin{equation}
\label{eq:susy}
H = \Sigma^2 - \frac12\,\sigma_3 = \left(
\begin{array}{cc}
 H_+  - \displaystyle \frac12 &  0   \\
 0     &  H_- + \displaystyle \frac12
\end{array}
\right)
\end{equation}
with
\begin{equation}
H_+ = a^-\,a^+\,, \qquad H_- = a^+\,a^-\,.
\end{equation}
This form of the harmonic oscillator Hamiltonian can be interpreted in terms of supersymmetric Quantum Mechanics (SUSY QM) 
\cite{GMR} because, as shown by Eq. \eqref{eq:JC} below, the operator $\Sigma$ can be understood as the sum of the supercharge 
operators associated with this specific problem. Let us remark that from the point of view of recovering the harmonic oscillator 
Hamiltonian, the expression \eqref{eq:Sigma} it's not the only one that can be adopted. However, it's easy to show that the general 
form
\begin{equation}
\tilde{\Sigma} = \frac1{\sqrt{\alpha^2 + \beta^2}}\,(\alpha\,\sigma_j + \beta\,\sigma_k) \qquad\qquad (j \neq k)\,,
\end{equation}
can always reduced to the form  \eqref{eq:Sigma} by means of a unitary transformation
\begin{equation}
\mathcal{U} = \exp\left\{i\,\sum_{m = 1}^3 a_m\,\sigma_m\right\}\,.
\end{equation}

The term proportional to $\sigma_3$ in Eq. \eqref{eq:susy} is linked to the vacuum field fluctuations. In the usual treatment of SUSY QM 
the ground state is associated only with the operator $H_-$, while the present formalism suggests that the vacuum field fluctuations can 
be interpreted as a contribution emerging from a kind of population inversion. We note indeed that, apart from its mathematical role, the 
operator $\Sigma$ is amenable for a transparent interpretation in physical terms. The joint use of the properties of Pauli matrices 
and ladder operators allows us to cast Eq. \eqref{eq:Sigma} in the form 
\begin{equation}
\label{eq:JC}
\Sigma = a^-\,\sigma_+ + a^+\,\sigma_-
\end{equation}
with
\begin{equation}
\sigma_+ = \frac12\,(\sigma_1 + i\,\sigma_2) = \left(
\begin{array}{cc}
  0  &  1   \\
  0  &   0
\end{array}
\right)\,, \qquad
\sigma_- = \frac12\,(\sigma_1 - i\,\sigma_2) = \left(
\begin{array}{cc}
  0  &  0   \\
  1  &   0
\end{array}
\right)\,.
\end{equation}
The operator \eqref{eq:JC} is formally equivalent to the interaction potential in Jaynes-Cummings Hamiltonian \cite{SchMuf}, describing 
the interaction of a quantized field with a two-level system, whose energies differ by the characteristic gap of the harmonic oscillator 
spectrum\footnote{This analogy is an a posteriori justification for our particular choice \eqref{eq:Sigma} for the operator $\Sigma$.}. 
We can therefore say that the use of the Dirac factorization method and the quantized nature of the Hamiltonian operator induces, 
in quite a natural way, an analogous quantized two level-structure. This, in turn, can be interpreted as the physical origin of the graded 
algebraic nature of the model we are developing.

The algebraic structure, underlying the Hamiltonian \eqref{eq:susy}, requires the embedding of bosonic ($a^\pm$) and fermionic (Pauli 
matrices) operators which close up to form a super-algebra. Following Refs. \cite{Witten,Cooper,deLima}, we introduce the operators
\begin{equation}
U_\pm = a^\pm\,\sigma_\pm \qquad\qquad V_\pm = a^\mp\,\sigma_\pm  
\end{equation}
that satisfy the commutation relations 
\begin{align}
\left[U_+, U_-\right] & = H\,\sigma_3 - \frac12  &\left[V_+, V_-\right]  =  H\,\sigma_3 + \frac12 \nn \\
\left[U_\pm, V_\pm \right] & = \mp\,\sigma_\pm^2  &\left[U_\pm, V_\mp\right] = \pm\,2\,K_\pm\,\sigma_3
\end{align}
where we have introduced the two operators 
\begin{equation}
K_\pm = \frac12\,(a^\pm)^2\,,
\end{equation}
that, together the Hamiltonian operator, generate the SU(1,1) algebra:
\begin{equation}
\left[H, K_\pm\right] = \pm\,2\,K_\pm \qquad\qquad \left[K_+, K_-\right] = - H
\end{equation}
and allows us to recover the full structure of the ortho-symplectic algebra $\mathfrak{osp}(1|2)$ with the remaining commutation 
brackets (the not mentioned brackets are zero)
\begin{equation}
\left[K_\pm\, U_\mp\right] = \mp\,V_\mp \qquad\qquad \left[K_\mp, V_\pm\right] = \mp\,U_\pm\,.
\end{equation}

As anticipated, the operators $V_\pm$ play the role of supercharges of the system. Indeed, from Eq. \eqref{eq:JC}, we can 
write $\Sigma = V_+ + V_-$, and, since $V_\pm^2 = 0$, the supersymmetric part of the Hamiltonian is given by 
$\Sigma^2 = \{V_+, V_-\}$.

By using the previous identities, it's easy to show that the Heisenberg equation of motion for the operators $\sigma_k$ 
are:
\begin{equation}
\label{eq:bloch}
\dot{\sigma}_1 = \dot{\sigma}_2 = 0 \qquad \qquad \dot{\sigma}_3 = 2\,\{q, p\} = 4\,i\,(K_+ - K_-)
\end{equation}
from which it's easy to show that
\begin{equation}
\sigma_3 (t) \propto \sin^2 (\sqrt{2}\,t)\,,
\end{equation}
that represents the vacuum field spontaneous emission. As for the time evolution of the operator $\Sigma$ we obtain
\begin{equation}
\ddot{\Sigma} = - \Sigma
\end{equation}
i.e., a simple harmonic oscillation, physically associated with the emission and the absorption of a photon.

We have so far shown that the Dirac factorization method ÒdressesÓ the harmonic oscillator with a spin-like (two-level structure), 
yielding a graded Lie algebraic structure. It is therefore natural to ask about the real physical meaning of such a superimposed 
structure. The main representative of the fermionic structure is the vector $\vec{\sigma} = (\sigma_1, \sigma_2, \sigma_3)$, 
which may be considered as rather artificial. However, apart from the fact that it emerges in a very natural way from the mathematical 
procedure, its role is physically understandable and should be interpreted as that of a quantized two-level structure induced by the 
noncommuting nature of the $q$ and $p$ variables. More in general, we can view the vacuum as an ensemble of coupled 
two-level systems continuously emitting and absorbing a photon.

Before closing the section, we briefly show how the Dirac factorization proceeds in the case of the slightly more complicated 
Hamiltonian
\begin{equation}
\label{eq:hamf}
H = \frac12\,\left[p^2 + f (q)\right]\,.
\end{equation}
If, in analogy with Eq. \eqref{eq:Sigma}, we introduce the operator
\begin{equation}
\label{eq:ups}
\Upsilon = \frac1{\sqrt{2}}\,(\sqrt{f (q)}\,\sigma_1 - p\,\sigma_2) = \left(
\begin{array}{cc}
       0  &  A^-  \\
   A^+  & 0   
\end{array}
\right)
\end{equation}
with 
\begin{equation}
\label{eq:ladd}
A^\pm = \frac1{\sqrt{2}}\,(\sqrt{f (q)} \mp i\,p)\,,
\end{equation}
we can write
\begin{equation}
H = \Upsilon^2 - \frac14\,\frac{f^\prime (q)}{\sqrt{f (q)}}\,\sigma_3 = \left(
\begin{array}{cc}
 A^-\,A^+  - \displaystyle \frac14\,\frac{f^\prime (q)}{\sqrt{f (q)}} &  0   \\
 0     &  A^+\,A^- +\displaystyle \frac14\,\frac{f^\prime (q)}{\sqrt{f (q)}}
\end{array}
\right)\,.
\end{equation}
Also in this case
\begin{equation}
\mathcal{V}_- = \left(
\begin{array}{cc}
 0 & A^-   \\
 0 & 0
\end{array}
\right) \qquad\qquad 
\mathcal{V}_+ = \left(
\begin{array}{cc}
 0     &  0 \\
 A^+ & 0
\end{array}
\right)
\end{equation}
are recognized as super-charges associated with the Hamiltonian \eqref{eq:hamf}, and the quantities
\begin{equation}
f_\pm = f (q) \pm \frac14\,\frac{f^\prime (q)}{\sqrt{f (q)}}
\end{equation}
are interpreted as super-partners potentials. The example which follows has a twofold motivation, i.e., an explicit application 
of the techniques we have developed and the use of nonstandard special functions emerging from the analysis of the problem under 
study. We consider the case of the quartic oscillator, i.e. $f (q) = \lambda\,q^4$, for which the ladder operators are
\begin{equation}
A^\pm = \frac1{\sqrt{2}}\,(\sqrt{\lambda}\,q^2 \mp i\,p)
\end{equation}
with
\begin{equation}
[A^-, A^+] = 2\,\sqrt{\lambda}\,q\,.
\end{equation}
The states can be defined as follows
\begin{equation}
\label{eq:phin}
\varphi_n = \frac1{\sqrt{n!}}\,(A^+)^n\, | 0 \rangle
\end{equation}
where, from the condition $A^-\,| 0 \rangle = 0$, for the vacuum it is possibile to deduce the following expression
\begin{equation}
| 0 \rangle \propto \exp\left(- \frac{\sqrt{\lambda}}3\,q^3\right) \qquad\qquad (q > 0)\,.
\end{equation}
The explicit form of the functions \eqref{eq:phin} can be obtained by using the identities (see Appendix) 
\begin{align}
(A^+)^n & = \frac1{\sqrt{2^n}}\,\sum_{k = 0}^n \binom{n}{k}\,(- 1)^k\,H_{n - k}^{(3)} 
\left(\sqrt{\lambda}\,q^2, \sqrt{\lambda}\,q, \frac{\sqrt{\lambda}}3\right)\,\partial_q^k \\
\partial_q^k e^{- \sqrt{\lambda}\,q^3/3} & = H_k^{(3)} \left(- \sqrt{\lambda}\,q^2, - \sqrt{\lambda}\,q, - \frac{\sqrt{\lambda}}3\right)\,
e^{- \sqrt{\lambda}\,q^3/3} \label{eq:dex3}
\end{align}
where  $H_n^{(3)}$ are the third-order Hermite polynomials whose definition is given in Appendix. The use of the addition theorem  
\begin{equation}
\sum_{k = 0}^n \binom{n}{k} H_{n - k}^{(3)} (x_1, x_2, x_3)\,H_k^{(3)} (y_1, y_2, y_3) = H_n^{(3)} (x_1 + y_1, x_2 + y_2, x_3 + y_3)
\end{equation}
and taking into account that $(- 1)^k\,H_k^{(3)} (x, y, z) = H_k^{(3)} (- x, y, -z)$, finally yields 
\begin{equation}
\varphi_n \propto \frac1{\sqrt{2^n\,n!}}\,H_n^{(3)} \left(2\,\sqrt{\lambda}\,q^2, 0, \displaystyle \frac23\,{\sqrt{\lambda}}\right)\,
e^{- \sqrt{\lambda}\,q^3/3}\,.
\end{equation}
The functions $\varphi_n$ are significantly different from the ordinary harmonic oscillator functions, and the discussion of their properties 
is out of the scope of this paper. 

In the forthcoming section we will consider more physical examples yielding further elements supporting the physical reality of 
the mathematical devices discussed so far.

\section{Physical examples}
We have mentioned that the quantized nature of the spin-like structure introduced in the previous sections reflects the quantum 
nature of $q$ and $p$ variables. The same happens with the so-called Landau states, emerging in the quantum 
analysis of the motion of a particle with electric charge $e$ in a classical magnetic field of intensity $B$ \cite{Cohen}.

By identifying the $z$-axis of the reference frame with the direction of the field, the associated vector potential writes 
$\vec{A} = (0, B\,x, 0)$. This choice is not unique but it ensures that the magnetic field is orthogonal to the plane of motion, 
i.e., $\vec{p} = (p_x, p_y, 0)$. The relativistic Hamiltonian operator ruling such a process can be written as\footnote{In 
this section we restore the mass $m$ of the particle and the constants $\hbar$ and $c$.}
\begin{align}
H &= c\,\sqrt{\left(\vec{p} - e\,\vec{A}\right)^2 + (m\,c)^2} \nn \\
    &= c\,\sqrt{p_x^2 + p_y^2 - 2\,e\,B\,x\,p_y + (e\,B\,x)^2 + (m\,c)^2}\,.
\end{align}
Since $[p_y, H] = 0$, the operator $p_y$ can be replaced by its eigenvalue $\hbar\,k_y$, and the Hamiltonian can be rewritten 
as 
\begin{equation}
H = c\,\sqrt{p_x^2 + m^2\,\omega_c^2\,X^2 + (\hbar\,k_y)^2\,-\,m\,\omega_c^2\,x_B^2 + (m\,c)^2}\,.
\end{equation}
with
\begin{equation}
\omega_c = \frac{| e |\,B}{m\,c}\,, \qquad x_B = \frac{\hbar\,k_y}{m\,\omega_c}\,, \qquad X = x - x_B\,. \nn
\end{equation}
By introducing the operator 
\begin{equation}
W =  p_x\,\sigma_1 + m\,\omega_c\,X\,\sigma_2 + \sqrt{(\hbar\,k_y)^2\,-\,m\,\omega_c^2\,x_B^2 + 
(m\,c)^2}\,\sigma_3
\end{equation}
it's easy to show that
\begin{equation}
\label{eq:landau}
H = c\,\sqrt{W^2 - m\,\hbar\,\omega_c\,\sigma_3}\,.
\end{equation}
In the common experimental situations, the term proportional to $\sigma_3$ can be neglected, since, for values of the magnetic field 
around 0.1 T (or even larger) it is of the order of meV, while the term $c\,W$ varies in the region of hundreds of keV. Under this 
approximation, the study of the relativistic Landau states reduces to a Jaynes-Cummings problem. The Hamiltonian \eqref{eq:landau} 
describes the dynamics of a two-level system with a level spacing fixed by the strength of the magnetic field. The supersymmetric nature 
of the Jaynes-Cummings model has been discussed elsewhere \cite{SchMuf} and we will not dwell on it. Here we note that a physical 
realization of such a system is given, for example, by a Free Electron Laser (FEL) source where a relativistic beam of electrons propagates 
inside an axial magnetic field \cite{FEL}. The possibility of using such a point of view to construct a laser-like theory for the FEL-like devices 
has been partially considered in \cite{ElgFul}, and will be the topic of a future speculation.

Before closing this section we will reconsider a point partially touched in Ref. \cite{PRA}, and in previous sections, concerning 
the evolution of the spin-like system associated with the harmonic oscillator via the Dirac factorization method. To simplify the 
problem we will refer to the relativistic 1-dimensional Hamiltonian
\begin{equation}
H = c\,\sqrt{p^2 + (m\,c)^2}
\end{equation}
that, for example, can be factorized as follows
\begin{equation}
\label{eq:1dham}
H = c\,(p\,\sigma_1 + m\,c\,\sigma_3)
\end{equation}
The physical meaning of the above Hamiltonian has been discussed in Ref. \cite{PRA}, where it has been stressed that it should 
not be confused with the Dirac Hamiltonian nor with the Pauli counterpart. Here, we will use it as a toy model to get a further 
support to the previous speculations.

The equations of motion for the vector $\vec{\sigma}$ are
\begin{equation}
\frac{\de}{\de t}\,\vec{\sigma} = \vec{\Omega}\,\times\,\vec{\sigma} \qquad\qquad \vec{\Omega} = \frac1{\hbar}\,(p\,c, 0, m\,c^2)\,.
\end{equation}
This equation describes a purely quantum motion (with no classical counterpart) which should be understood as a kind of 
\textit{zitterbewegung} \cite{Barut,PRA} (a trembling motion due to the interference between negative and positive states 
contained in the Hamiltonian \eqref{eq:1dham}). Even though there is not any explicit presence of ladder operators, the 
Hamiltonian \eqref{eq:1dham} contains a hidden two-level structure and a supersymmetric underlying algebra. In fact, by 
expressing the momentum in terms of the ladder operators, we get ($g = c/\sqrt{2\,\hbar}$)
\begin{equation}
H = m\,c^2\,\sigma_3 + i\,\hbar\,g\,(a^+ + a^-)\,(\sigma_+ - \sigma_-)\,, 
\end{equation}
which writes as a Jaynes-Cummings Hamiltonian, but without the \textit{rotating wave approximation} assumption.
 
\section{Concluding remarks}\label{s:conrem}
As it is shown in sec. \ref{s:dirac}, the method of factorization exhibits, in a fairly natural way, all the essential features of SUSY QM: 
i) the vacuum field energy is factorized out; ii) two potentials with isospectral properties are recovered. But, as often happens, 
nothing is really new and the procedures leading to super-potentials traces back to methods known well before to the birth of quantum 
mechanics itself \cite{Dattoli,Millson}\footnote{In particular, in Ref. \cite{Dattoli} the problem has been treated by showing that a 
second-order differential equation with non-constant coefficients can be written in terms of bi-orthogonal partner solutions, a posteriori 
recognized as supersymmetric components.}. In order to appreciate this point, let us consider the following second-order differential 
equation:
\begin{equation}
\label{eq:zetax}
z^{\prime\prime} (x) + \mu_- (x)\,z (x) = 0\,.    
\end{equation}
If the ``potential" $\mu_- (x)$ is expressible as
\begin{equation}
\mu_- (x) = - \frac14\,\phi^2 (x) - \frac12\,\phi^\prime (x)\,,
\end{equation}
i.e., is solution of a Riccati equation, the solution of Eq. \eqref{eq:zetax} can written as 
\begin{equation}
z (x) = \exp\left\{\frac12\,\int^x \de \xi\,\phi (\xi)\right\}\,u_- (x)
\end{equation}
with the function $u_- (x)$ satisfying the differential equation 
\begin{equation}
u^{\prime\prime}_ - (x) + \phi (x)\,u^\prime_ - (x) = 0\,.
\end{equation}
Therefore we get 
\begin{equation}
z (x) = \exp\left\{\frac12\,\int^x \de \xi\,\phi (\xi)\right\}\,\int^x \de \eta\,\exp\left\{- \int^\eta \de \nu\,\phi (\nu)\right\}\,.
\end{equation}
Furthermore once $\phi (x)$ is fixed, we can define a second ``potential"
\begin{equation}
\mu_+ (x) = - \frac14\,\phi^2 (x) + \frac12\,\phi^\prime (x) 
\end{equation}
(the super-partner of $\mu_- (x)$, according to the present terminology) which specifies the solutions of a second differential 
equation,  obtained from the first by just replacing $\phi (x)$ with $- \phi (x)$. 

The outlined procedure is essentially the Liouville method to reduce a second-order differential equation to its standard 
form\footnote{We remind that, given the differential equation 
$y^{\prime\prime} + a (x)\,y^\prime (x) + b (x)\,y (x) = 0$, the Liouville transformation
$$
y (x) = \exp\left\{- \frac12\,\int^x \de \xi\,a (\xi)\right\}\,v (x)
$$
reduces the initial equation to $v^{\prime\prime} (x) + c (x)\,v (x) = 0$, 
where $c (x) = b (x) - \frac14\,[a^2 (x) + 2\,a^\prime (x)]$.}.
However, it is remarkable that all the features concerning the supersymmetry are recovered in a very natural way by just 
applying the Dirac factorization method, which leads to the two-level Hamiltonian \eqref{eq:hamf} with the role of the vacuum 
field factorized out from the very beginning. 

Let us come back to the operator $\Upsilon$ defined in Eq. \eqref{eq:ups}. Also in this case, we can write
\begin{equation}
\Upsilon = A^-\,\sigma_+ + A^+\,\sigma_-\,,
\end{equation}
and, therefore, it can be be interpreted as the interaction between a two-level system and the bosonic field defined by ladder 
operators given in Eq. \eqref{eq:ladd}.

More in general, we can conjecture that the Jaynes-Cummings model can be generalized by the following interaction Hamiltonian
\begin{equation}
\label{eq:genJC}
H_{\mathrm{JC}} = \Upsilon - \frac12\,\omega (q)\,\sigma_3 \qquad\qquad 
\left(\omega (q) \propto \frac{f^\prime (q)}{\sqrt{f (q)}}\right)\,.
\end{equation}
As a consequence of the more complicated commutation relations involved, the algebraic nature of this operator is less direct than 
in the case of the ordinary J-C operator. However, the study of quantum states ruled by the interaction Hamiltonian \eqref{eq:genJC} can 
be done using the evolution operator associated with the Hamiltonian The properties of this operator can be discussed by using the 
methods developed in the past to treat evolution problems in quantum mechanics (see Refs. \cite{Fujii1,Fujii2}). For example, using the 
so-called symmetric-split decomposition method \cite{BandShe}, the evolution operator can be approximated as follows 
\begin{equation}
\label{eq:evol}
\Psi (t) = \prod_{j = 1}^N U_j\,\Psi (0)
\end{equation}
with 
\begin{equation}
U_j = e^{- i\,t_j\,\Upsilon/2}\,e^{- i\,t_j\,\omega (q)\,\sigma_3}\,e^{- i\,t_j\,\Upsilon/2} + O (t_j^3)
\end{equation}
The use of standard identities for exponential operator (see Ref. \cite{Fujii2}), along with the iterated application of the evolution operator 
in Eq. \eqref{eq:evol}, yields an efficient way of calculating the evolution of these quantum states.

In this paper we have touched different topics, whose underlying leitmotif is the Dirac factorization. We have discussed the relativistic quantum 
mechanics, the zitterbewegung, the Jaynes-Cummings model, the relativistic Landau levels, and the Dirac oscillator (even though not explicitly 
mentioned). Most of these effects are difficult to be subjected to experimental investigation. For example, the electron zitterbewegung 
exhibits oscillations at very large frequencies ($\sim10^{21}$ Hz), not accessible to currently available experimental techniques. Such drawback 
is overcome in the context referred to as quantum simulation \cite{Gerri}, where the trembling motion may occur, for example, in crystalline 
solids when their band structure is represented by a two-band model reminiscent of the one-dimensional Dirac equation \cite{Cannat}. The 
explicit realization of the simulation \cite{Gerri, Dreisow} has provided strong indications for the existence of such a genuine quantum behavior. 
Further experiments using quantum simulator techniques are suggested to test other quantum paradoxes like the Klein paradox \cite{Longhi}.

We believe that the topics treated in this paper can be framed within the context of the quantum phenomenology which can be experimentally tested 
using quantum simulators. A more specific analysis in this direction will be developed elsewhere.

\appendix

\section{}
In section \ref{s:conrem} we have used the higher-order Hermite polynomials to express the ordered form of the 
operator $O_n = \left(\partial_x + P (x)\right)^n$. These polynomials are defined through the generating function
\begin{equation}
\label{eq:Aherm}
\sum_{n = 0}^\infty \frac{t^n}{n!}\,H_n^{(m)} (x_1, x_2, \cdots, x_m) = \exp\left\{\sum_{k = 1}^m x_k\,t^k\right\}\,.
\end{equation}
and can be constructed recursively according to formula
\begin{equation}
H_n^{(m)} (x_1, x_2, \cdots, x_m) = n!\,\sum_{k = 0}^{[n/m]} 
\frac{x_m^k}{k!\,(n - m\,k)!}\,H_{n - m\,k}^{(m - 1)} (x_1, x_2, \cdots, x_{m - 1})\,.
\end{equation}
By using the generating function methods, we can define the following operator
\begin{equation}
E (x, t) = \sum_{n = 0}^\infty \frac{t^n}{n!}\,O_n = \exp\{t\,[\partial_x + P (x)]\}
\end{equation}
that satisfies the differential equation
\begin{equation}
\partial_t\,E (x, t) = [\partial_x + P (x)]\,E (x, t) \qquad\qquad E (x, 0) = 1
\end{equation}
whose solution is  
\begin{equation}
E (x, t) = \exp\left\{t\,\partial_x + \int_0^t \de t^\prime\,P (x - t^\prime)\right\} = 
\exp\left\{\int_0^t \de t^\prime\,P (x + t - t^\prime)\right\}\,e^{t\,\partial_x}\,.
\end{equation}

In the case $P (x) = \alpha\,x^2$, performing the integral and using the Eq. \eqref{eq:Aherm}, we easily 
obtain 
\begin{equation}
E (x, t) = \sum_{k = 0}^\infty \frac{t^k}{k!}\,H_k^{(3)} \left(\alpha\,x^2, \alpha\,x, \frac{\alpha}3\right)
\,e^{t\,\partial_x}
\end{equation}
and, thus, 
\begin{equation}
O_n = \sum_{k = 0}^n \binom{n}{k}\,H_{n - k}^{(3)} \left(\alpha\,x^2, \alpha\,x, \frac{\alpha}3\right)
\,\partial_x^{\,k}\,.
\end{equation}

As for the successive derivatives of the function $e^{- x^3}$, by applying the generating function method 
we get 
\begin{equation}
\sum_{n = 0}^\infty \frac{t^n}{n!}\,\partial_x^n\,e^{- x^3} = e^{t\,\partial_x}\,e^{- x^3} = e^{- (x + t)^3} = 
\sum_{n}^\infty \frac{t^n}{n!}\,H_n^{(3)} (- 3\,x^2, - 3\,x, - 1)\,e^{- x^3}
\end{equation}
and, therefore, from the comparison of the same powers of $t$, the second identity in \eqref{eq:dex3}. 
%
%
%
%

\end{document}